\documentclass[letterpaper, 10 pt, conference]{ieeeconf}  
\usepackage{amsmath}
\usepackage{graphicx}
\usepackage{xcolor}
\usepackage{tikz}
\usepackage{circuitikz}
\usetikzlibrary{positioning}
\usetikzlibrary{shapes,arrows}
\usepackage{caption}
\usepackage{subcaption}
\usepackage{amsthm}
\usepackage{amssymb}
\usepackage{wrapfig}

\newtheorem{proposition_1}{\bf{Proposition}}

\IEEEoverridecommandlockouts                              

\title{\LARGE \bf
Control Designs for Critical-Continegency Responsible Grid-Following Inverters and Seamless Transitions To and From  Grid-Forming Modes
}

\author{Jaesang Park\textsuperscript{1,a}, Alireza Askarian\textsuperscript{1,b}, and Srinivasa Salapaka\textsuperscript{1,c}
\thanks{\textsuperscript{1} Department of Mechanical Science and Engineering, University of Illinois at Urbana-Champaign, 61801 IL, USA
\textsuperscript{a}jaesang4@illinois.edu,\textsuperscript{b}askaria2@illinois.edu,\textsuperscript{c}salapaka@illinois.edu}%
}

\begin{document}

\maketitle
\thispagestyle{empty}
\pagestyle{empty}

\begin{abstract}

This article introduces two control frameworks: one for Grid-Following (GFL) inverters aiding Grid-Forming (GFM) inverters in voltage regulation during large contingency events and optimizing power transactions under normal conditions; and another for seamless transitions between grid-tied and grid-isolated setups, managing voltage transient characteristics. In microgrids, GFM inverters regulate voltage, while GFL inverters handle power transactions. The proposed GFL control detects abrupt load/generation changes, adjusting power transactions using local storage to support GFM inverters during contingencies. Additionally, a transition control ensures smooth GFL-GFM shifts, reducing power and voltage fluctuations. Simulation results validate improved voltage regulation during contingencies and enhanced power tracking during slow changes, alongside minimized transient overshoot.

\end{abstract}




\section{INTRODUCTION}

With increasing concern for climate change, global efforts have been made in recent years to reduce carbon emissions. To mitigate the pollution from fossil-based power generation and achieve lower carbon emissions, renewable energy sources (RES) such as solar, wind, and hydro are being integrated on a large scale into the power network. 
These RES, often supplemented with energy storage devices like batteries, are connected to the grid through DC/AC inverters. Currently, most inverters operate in grid-following (GFL) mode, where the grid controls voltage and frequency while the inverter supplies power. However, achieving high RES penetration with GFL inverters poses challenges.

In microgrids, especially when grid-isolated, sudden large power events like load or generational changes can strain voltage regulation, resulting in significant deviations from nominal values. Similar deviations happen in grid-tied setups due to factors such as weather disturbances and uncertainties in power production, which worsen with high penetration of RES. Conventional GFL-based inverter control designs lack robustness against these deviations, leaving grid viability solely on Grid-Forming (GFM) inverters. However, if GFM inverters are overwhelmed by sudden changes, cascading failures may lead to grid instability.

One explored approach \cite{elkhatib2017evaluation} is Frequency-Watt (FW) control. Here, the inverter maintains a desired power set-point within a specific frequency range. However, if the frequency exceeds this range, the inverter follows a droop curve instead of the precise set-point. While this method enhances grid stability, it struggles in weak microgrids where frequency control is lax, resulting in erratic power injection. Moreover, when the frequency drops below the nominal value, continuous power injection attempted by the inverter control design may not be sustainable over long intervals, posing operational challenges. In contrast, \cite{qi2021synthetic} introduces virtual inertia, simulating synchronous machine dynamics, where the energy stored in DC link capacitors to mechanical inertia in synchronous machines, aiming to emulate traditional power generation behaviors that enhance grid stability. While it excels in steady-state active power tracking and frequency support, it mainly focuses on the active power-to-frequency response, neglecting dynamics of reactive power and voltage.

Relevant studies, such as those by \cite{wang2022study} and \cite{chakraborty2023seamless}, explore control strategies transitioning GFL and GFM modes. Here inverter control is changed from GFL to GFM based on amplitude/frequency deviations from nominal values. While suitable for grid-tied operations with occasional deviations, in weaker microgrids, operating all inverters in GFM mode might not be efficient. Our proposed control design enables inverters to switch to GFM-like voltage regulation only during rapid load or generation changes, operating as GFL inverters during gradual changes. This assists full-time GFM inverters during extreme power fluctuations while providing commanded power, like maximum power point tracking, during stable periods.

We make use of algebraic structures in our control design to enable transitions between GFL and GFM modes; thus facilitating smooth transitions of microgrids between grid-tied and grid-isolated states. In the event of a main power grid anomaly, RES can disconnect and form a microgrid, preventing a complete blackout. Our transition capability ensures the grid-isolated microgrid includes a GFM inverter for regulating frequency and voltage, while allowing other GFL inverters to connect, ensuring the presence of a GFM inverter within the newly formed microgrid consistently. The proposed control retains the functional advantages and stability aspects of the control strategies developed in \cite{chakraborty2023seamless}. The main distinctive feature is that  our control design enables smooth transitions while {\em ensuring} safe transient voltage characteristics. In contrast, the strategy in \cite{chakraborty2023seamless} does not guarantee transient characteristics during transitions. For instance, their methods may introduce sudden jumps in the controller output when switching modes, potentially causing grid instability. 
Using the proposed GFL inverter, we demonstrate  about 44\% reduction in the Rate of Change of Frequency (RoCof) during sudden load changes compared to using a conventional GFL inverter. Additionally, with the proposed smooth transition algorithm, during the transition from GFL to GFM, we demonstrate that we can achieve only 1\% of overshoot in active power compared to the sudden transition and no overshoot in reactive power.
\section{Overview of The Conventional GFL and GFM Inverter} \label{RobustnessOfInverters}
\vspace{-5pt}
\subsection{Modeling of Inverter connected to a grid}

\begin{wrapfigure}{L}{0.30\textwidth}
\centering
\includegraphics[width=0.30\textwidth]{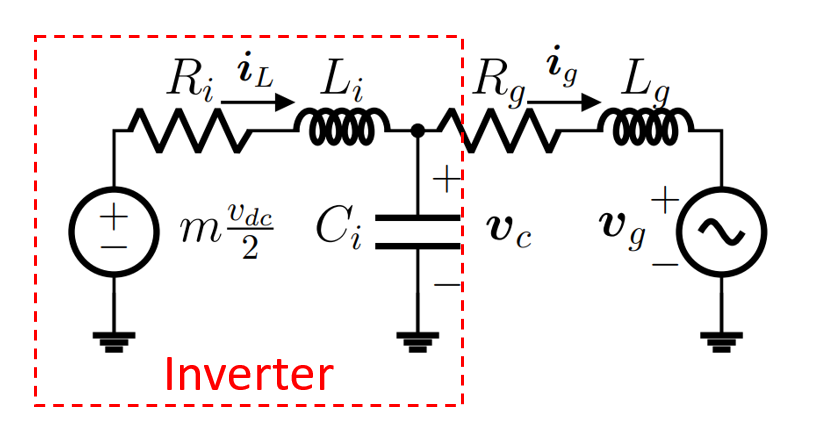}        
\caption{The model of an inverter with LC filter connected to a grid.}
\label{fig:circuits}
\vspace{-7pt}
\end{wrapfigure}

A DC source power (e.g., RES) is transacted with an AC power grid using an inverter. An inverter with an LC filter connected to a grid can be modeled as in Fig. \ref{fig:circuits}. It converts DC to AC using fast switching of transistors. A power converter (not shown here) takes power from the source and maintains a constant voltage $V_{dc}$ at its output. The control design on the converter achieves voltage regulation by using capacitance or battery storage on the DC side. The inverter's switch-averaged model, depicted in Fig. \ref{fig:circuits}, uses the modulating signal $m\in[-1,1]$ as a control parameter. This signal is actuated using pulse-width modulated (PWM) signals, which determine duty cycles governing the switching. The averaged model resembles a controlled voltage source with output voltage $m\frac{V_{dc}}{2}$. However, to suppress high-frequency switching noise, an LC filter with resistance $R_i$, inductance $L_i$, and capacitance $C_i$ is employed. This filter mitigates noise in the filter current $i_L$ and the capacitor voltage $V_c$, which serves as the inverter's output voltage interfacing with the grid. The grid, represented as a voltage source with voltage $V_g$, is separated from the inverter by line impedance, modeled by line resistance $R_g$ and inductance $L_g$. The output current $i_g$ from the inverter is the power transacted with the grid.

\subsection{Response of Conventional GFL to perturbation}

The main goal of a GFL is to control its output power. Since the output power depends on the output current $i_g$ and the capacitor voltage $V_c$, where $V_c$ can be measured, its goal is converted to a controlling output current $i_g$.

Note that typically the control signal $m$ is designed to make the inductor current $i_L$ track $i_{L,\text{ref}}$, where the reference $i_{L,\text{ref}}$ is appropriately designed to make $i_g$ track $i_{g,\text{ref}}$.  In direct-quadrature (dq) coordinates, the dynamics between $i_L$ and $i_g$ is given by the capacitor dynamics as follows \cite{yazdani2010voltage}
\begin{align}
    \begin{split}\label{eqn:ig_dym}
        &\hat i_g^{d,q}=\hat i_L^{d,q}-C_i\hat V_c^{d,q} s\pm C_i\omega \hat  V_c^{q,d}\\
    \end{split}
\end{align}
where superscript $d$ and $q$ respectively represent the direct and quadrature components, while the hat notation denotes the s-domain signal. Additionally, $\omega$ denotes the grid voltage frequency.

Here by choosing
$\hat i_{L,ref}^{d,q}=\hat i_{g,ref}^{d,q}\mp C\omega \hat V_c^{q,q}$ and designing the filter current controller with a large bandwidth,
the coupling terms $C\omega V_c^d$ and $C\omega V_c^q$ cancel out and $\hat i_{L}^{d,q}\approx \hat i_{L,ref}^{d,q}$ resulting in the following dynamics given by  
\begin{align}\label{eqn:GFL_IV_response}
\begin{split}
    &\hat V_c^{d,q}=\frac{1}{C_i s}(\hat i_{g,ref}^{d,q}-\hat i_{g}^{d,q})
\end{split}
\end{align}
In addition to this, the frequency of the GFL control is determined by the PLL, which takes $V_c^{q}$ as an input:
\begin{align}
\begin{split}\label{eqn:GFL_IW_response}
    \hat \omega
    &=K_{PLL}(s)\frac{1}{C_i s}(\hat i_{g,ref}^q-\hat i_g^q)+\hat \omega_0.
\end{split}
\end{align}
In this paper, we use subscript $0$ to denote nominal values and superscript $-$  to represent amplitude of the underlying signals. Note that by assuming that the PLL accurately estimates the phase of the capacitance voltage, one can easily deduce from $d$-$q$ transformations that  $V_c^d \approx {\bar V}_c$ and $V_c^q\approx 0$. Also, it is easy to show that 
$P=\frac{3}{2}(V_c^di_g^d+V_c^qi_g^q)$ and $Q=\frac{3}{2}(-V_c^di_g^q+V_c^di_g^q)$, $i_g^d$ and $i_g^q$ can be approximated as $i_g^d=\frac{2}{3V_0}$, $i_{g,ref}^d=\frac{2}{3V_0}P_0$, $i_g^q=-\frac{2}{3V_0}Q$, and $i_{g,ref}^q=-\frac{2}{3V_0}Q_0$.
See \cite{yazdani2010voltage} for these $d$-$q$ transformation details. 

From  (\ref{eqn:GFL_IV_response}) and (\ref{eqn:GFL_IW_response}), we obtain
\begin{align}\small
\begin{split}\label{eqn:Conv_GFL_inertia}
    &\hat{\bar{V_c}}=\frac{2}{3V_0 C_i s}(\hat P_0-\hat P),\quad\hat\omega=\hat\omega_0-\frac{2K_{PLL}(s)}{3V_0 C_i s}(\hat Q_0-\hat Q).
\end{split}
\end{align}
Since the PI controller is widely used for $K_{PLL}(s)$, without loss of generality, both equations in (\ref{eqn:Conv_GFL_inertia}) include at least one pure integrator. Therefore, they have infinite DC gain. Consequently, if there is a mismatch between the nominal output power and the actual output power, the GFL inverter cannot regulate voltage and frequency.

In addition, in (\ref{eqn:Conv_GFL_inertia}), the nominal voltage $V_0$ is a grid parameter that we cannot change. Moreover, $C_i$ is chosen to be small, just enough to reject the switching noise of a transistor, to minimize the effect of $V_c$ on $i_g$. This makes the bandwidth of $\frac{2}{3V_0C_i s}$ large. Therefore, a high-frequency disturbance in output power propagates to the inverter voltage and frequency, potentially leading to grid instability.

Some literature, like \cite{askarian2024enhanced}, proposes a GFL design that considers the bandwidth between active/reactive power and voltage/frequency. However, in most conventional designs, as mentioned earlier, this bandwidth is not adequately considered. Typically, only the bandwidth of the current controller and PLL are taken into account. Consequently, the bandwidth of (\ref{eqn:Conv_GFL_inertia}) unintentionally becomes large, as noted.

\subsection{Response of GFM to perturbation}
Unlike GFL, the goal of GFM is to control the output voltage $V_c$. The corresponding control designs typically accommodate uncertainties in power generation and consumption, which translate into mismatches between the commanded and actual power transactions.  Since physically it is impossible to accommodate these power mismatches without compromising the voltage (amplitude and frequency) regulation, the control laws allow the amplitude and frequency to {\it droop} or deviate from the nominal values; however, in a controlled manner. 

There are several types of droop control strategies depending on the types of output line impedance. However, even though the line impedance could vary in type, in most cases, it is inductive due to the presence of long transmission lines. Therefore, for simplicity, we assume a purely inductive line, and its droop strategy becomes the $Pf-QV$ type of droop \cite{zhong2016universal}, expressed as:
\begin{align}\label{eqn:PFQVdroop}\small
\begin{split}
    \hat{\omega}&=\hat{\omega_0}+K_P(\hat{P_0}-\hat{P}),\quad \hat{V}_{ref}=\hat{V_0}+K_Q(\hat{Q_0}-\hat{Q}).
\end{split}
\end{align}
Here, $K_P(s)=k_p\frac{\omega_{lpf}}{s+\omega_{lpf}}$, $K_Q(s)=k_q\frac{\omega_{lpf}}{s+\omega_{lpf}}$, and $k_p$ and $k_q$ are the slope of the curve, while $\frac{\omega_{lpf}}{s+\omega_{lpf}}$ is a low pass filter for power calculation.

Both dynamics given in (\ref{eqn:PFQVdroop}) have a finite dc-gain as they have no pole at the origin. Furthermore, $k_p$, $k_q$, and $\omega_{lpf}$ are design parameters, and by changing them, we can make the frequency and voltage insensitive to high-frequency disturbances of the output power. Therefore, even if there is a mismatch in nominal and actual output power, they can regulate the frequency and the magnitude of the voltage.

Therefore, when there are GFM inverters and conventional GFL inverters in the grid, when there is a disturbance of power, conventional GFL inverters do not contribute to voltage stability, and only GFM inverters respond to power mismatches and ensure a corresponding modified setpoint. However, in a network that has many GFL inverters, substantial and sudden power mismatches can overwhelm GFM inverters, thus compromising network stability. This motivates control designs where GFL inverters are more responsive to power uncertainties. 

\textbf{Remark:} The power-grid network robustness discussed above is also referred to as inertia in some existing literature such as  \cite{qi2021synthetic} and \cite{driesen2008virtual}, where intuition is developed by developing analogies with a synchronous generator used in well-studied power plants. Here, we establish this connection in the context of our analysis. In a synchronous machine, the mechanical rotor acts as an energy storage device, the energy stored in the rotor being given by $E=\frac{1}{2}J\omega^2$, where $J$ is the rotor moment of inertia, and $\omega$ is the rotating speed. Therefore, the rotor provides a mechanical means of absorbing or providing instantaneous power and prevents sudden changes in the rotating frequency of a synchronous machine. The swing equation and governor equations in (\ref{eqn:Govornor}) capture the dynamical behavior of the rotor frequency
\begin{align}
\begin{split}
    J\frac{d\omega}{dt} = T_{in}-T_{out}-D\omega,\quad
    T_{in} = T_0+k_\omega(\omega_0-\omega),
    \label{eqn:Govornor}
\end{split}
\end{align}
where $T_{in}$ is input torque comes from the governor, $T_{out}$ is torque output due to electricity generation, and $T_0$ is nominal torque. Furthermore, $D$ represents the damping of the rotor and $k_\omega$ is the governor droop coefficient. Using the approximation $P_0=T_0\omega_0$ and $P=T_{out}\omega_0$, where $P$ and $P_0$ are, respectively, the actual and nominal output power. Combining (\ref{eqn:Govornor}) together, as shown in \cite{driesen2008virtual}, the dynamics of a synchronous generator can be expressed as
\begin{align}
     \hat{\omega}=\hat{\omega}_0
     + \frac{1/H}{s+(k_\omega+D)/H}(\hat{P}_0 -\hat{P}),
     \label{eqn:VSG}
\end{align}
where $H=J/\omega_0$. As shown in \cite{liu2015comparison}, by defining $\omega_{lpf}=(k_\omega+D)/J$ and $k_p=1/(k_\omega + D)$, we can recover the first equation of (\ref{eqn:PFQVdroop}). Also, as the mechanical inertia $J$ increases, the dynamics are less sensitive to changes in power. Therefore, inertia has an equivalent meaning in this context.
\section{Proposed GFL controller}
We propose a control design in which the inverter responds to sudden and fast changes in power mismatches while being insensitive to slow changes. 
Here, we use our observations from the GFM droop-control design in the previous section, where we propose appropriately designing filters in the droop-control design to enable controllers that are sensitive to time constants of the load/generation perturbations.  To be more specific, we will design controllers that mimic conventional GFL designs at low frequencies while mimicking the dynamics of (\ref{eqn:PFQVdroop}) at frequencies higher than its cut-off frequency $\omega_{lpf}$.
\subsection{Power-injecting dynamics}
\begin{figure}[h]
    \centering
        \includegraphics[width=0.43\textwidth]{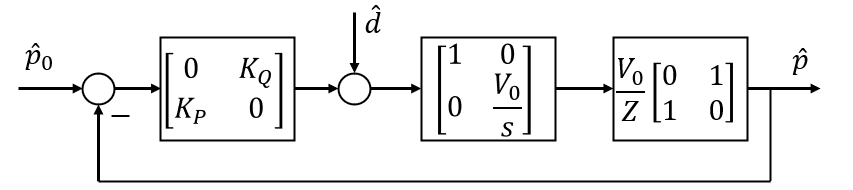}
    \caption{Block diagram of droop-based power injecting dynamics}
    \label{fig:power_block_diagram}
    \vspace{-5pt}
\end{figure}

The primary objective of GFL inverter is to control its output power. Designing such a controller using a droop-control structure requires analyzing the power-injecting dynamics. The droop equation relates output voltage amplitude and frequency to nominal and actual active/reactive power. However, these voltages and frequencies also impact active/reactive power. Illustrated in \cite{askarian2021control}, this voltage-frequency-to-power relationship forms a feedback loop. Analyzing closed-loop power injection dynamics necessitates understanding power transaction dynamics. This relationship is given  in the context of Fig \ref{fig:circuits}, where the power transactions ($P$ and $Q$) between the inverter output (specified by the output of the capacitor) and the point of coupling at the grid are given by 
$P+jQ=\bar{V_c}e^{j\theta_c}\big(\frac{\bar{V_c}e^{j\theta_c}-\bar{V_g}e^{j\theta_g}}{\bar{Z_g}e^{j\phi}}\big)^*,$
where $\bar{Z_g} e^{j \phi}=R_g+j\omega_0 L_g$ represents the line impedance.
Assuming a long transmission line, we have $\phi\approx 90^\circ$. By linearizing this equation around $\theta_c-\theta_g=0$ and $\bar V_g=V_0$, we obtain
$\begin{bmatrix}P\\Q\end{bmatrix}=\frac{V_0}{\bar{Z_g}}\begin{bmatrix}0&1\\1&0\end{bmatrix}\begin{bmatrix}\bar{V_c}-\bar{V_g}\\V_0 \delta\end{bmatrix}.$
In what follows, we assume that we have a voltage controller with a large bandwidth and therefore $\hat V_c=\hat V_{c,ref}$. Now we relate $\omega$ to $\delta$ by $\delta=\int_{0}^t (\omega-\omega_0) ds-\int_{0}^t(\omega_g-\omega_0) d\tau.$
By combining (\ref{eqn:PFQVdroop}) and  using these relations, we derive the block diagram of closed-loop power injecting dynamics as shown in Fig. \ref{fig:power_block_diagram}. Here, $\hat d= [\hat{\bar{V_g}}-\hat{\bar{ V_0}},\hat \omega_g-\hat \omega_0]^T$ and $\hat p= [\hat P,\hat Q]^T$ respectively represent amplitude and frequency deviations, and actual active and reactive power components.

Since the dynamics from $\hat P_0$ to $\hat P$ and the dynamics from $\hat Q_0$ to $\hat Q$ are decoupled, they can be expressed as two Single-Input Single-Output (SISO) systems as
\begin{align}
    \hat P&=\frac{V_0^2K_P}{s\bar{Z_g}+V_0^2K_P} \hat P_0+\frac{V_0^2}{s\bar{Z_g}+V_0^2K_P} (\hat \omega_g-\hat \omega_0),\label{eqn:Active_power_tf}\\
    \hat Q&=\frac{V_0K_Q}{\bar{Z_g}+V_0K_Q}\hat Q_0+\frac{V_0}{\bar{Z_g}+V_0K_Q}(\hat{\bar{V_g}}-\hat{\bar{ V_0}}).\label{eqn:Reactive_power_tf}
\end{align}

\subsection{Design of the controller}
For our control design, we consider the following criteria
\subsubsection{Reference tracking}
First, as discussed in the previous section, the goal of the GFL inverter is power tracking. At the very least, this should achieve steady-state power tracking. In frequency domain, it is is equivalent to $\frac{\hat p(j0)}{\hat p_0(j0)}\approx I$ and $\frac{\hat p(j0)}{d(j0)}\approx 0$. To achieve this, we propose the following control structure for $[K_P,K_Q]=:$
\begin{eqnarray}\label{eqn:Proposed_cont1} \left[\big(k_{p,P}+\frac{k_{i,P}}{s+\epsilon_P}\big), \big(k_{p,Q}+\frac{k_{i,Q}}{s+\epsilon_Q}\big)\right]\left(\frac{\omega_{lpf}}{s+\omega_{lpf}}\right).&&\end{eqnarray}
Note that setting $\epsilon_P=\epsilon_Q=0$ achieves $\frac{h(j0)}{h_0(j0)}=I$ and $\frac{h(j0)}{d(j0)}=0$. Additionally, by choosing $\epsilon_P$, $\epsilon_Q>0$, one can introduce robustness as a trade-off between performance. Finally, $\epsilon_P$ and $\epsilon_Q$ will be used in transition between GFL and GFM, a topic we will discuss later.
\subsubsection{Stability}

With the controller given in (\ref{eqn:Proposed_cont1}), the losed-loop characteristic polynomial for the system in (\ref{eqn:Active_power_tf}) is 
    $s^3 +(\omega_{lpf}+\epsilon_P)s^2+(\omega_{lpf}\epsilon_P+\frac{V_0^2}{Z_g}\omega_{lpf}k_{p,P})s
    +\frac{V_0^2}{Z_g}\omega_{lpf}(k_{p,P}\epsilon_P+k_{i,P})$.
Then, from the Routh-Hurwitz stability criterion if $\omega_{lpf}-k_{i,P}/k_{p,P}>0$, it is stable. Also, for the second-order system in (\ref{eqn:Reactive_power_tf}), since all the coefficients are positive, it stable for any $k_{p,Q}, k_{i,Q}>0$.

\subsubsection{Robustness against a high-frequency perturbation}
Finally, among the controllers that meet the reference tracking and stability conditions mentioned above, our objective is to find the controller that mimics the response in (\ref{eqn:PFQVdroop}) for $\omega>\omega_{lpf}$. With small $\epsilon_P$, $\epsilon_Q$, this can be achieved by setting $k_{i,P}=k_p$ and $k_{i,Q}=k_q$ while ensuring that $k_{i,P}/k_{p,P}\leq\omega_{lpf}$ and $k_{i,Q}/k_{p,Q}\leq\omega_{lpf}$.


\section{Smooth Transition between GFL and GFM}
Now we consider the implementation of control schemes that achieve transitions from GFL to GFM modes or vice versa, as desired. These transitions may be motivated by various reasons - for instance, decisions to go from grid-tied to grid-isolated modes. It turns out that we can achieve this by modifying parameters $\epsilon_P$ and $\epsilon_Q$ of our control design in (\ref{eqn:Proposed_cont1}). Specifically, when $\epsilon_P,\;\epsilon_Q$ small, we obtain the GFL controller as previously designed. Conversely, selecting $\epsilon_P=\epsilon_Q=\epsilon_{\text{max}}$ for some large $\epsilon_{max}$ mimics a GFM defined in (\ref{eqn:PFQVdroop}), as it derives $\frac{k_{i,P}}{s+\epsilon_P}\approx 0$ and $\frac{k_{i,Q}}{s+\epsilon_Q}\approx 0$.

\subsection{Proposed transition strategy}
\subsubsection{GFL to GFM}
The transition from GFL to GFM proposed in \cite{chakraborty2023seamless} is equivalent to change $\epsilon_P$ and $\epsilon_Q$ in (\ref{eqn:Proposed_cont1}) from 0 to $\infty$ abruptly. However, this abrupt change introduces a sudden variation in the control signal, potentially causing high-frequency perturbations in the grid and leading to grid instability. 
To mitigate this issue, we propose a smooth transition method that gradually ramps up $\epsilon_P$ and $\epsilon_Q$ from 0 to $\epsilon_{max}$. 
\subsubsection{GFM to GFL}
On the other hand, when switching from GFM to GFL, even if we suddenly make $\epsilon$ to 0, the control signal does not experience abrupt changes. Therefore, to facilitate the transition from GFM to GFL, we use a jump from $\epsilon_{max}$ to 0.
\subsection{Transition stability}
First consider the transition stability of active power dynamics during the transition. As shown in the previous section, if a controller satisfies the stability conditions mentioned above, 
the system in (\ref{eqn:Active_power_tf}) is stable for each $\epsilon_P\in[0,\epsilon_{max}]$. However, when $\epsilon_P$ varies, it becomes non-autonomous system and the stability of each time instance is not sufficient to guarantee overall stability. 

Now, let $A_P(\epsilon_P)$ be the state matrix of a state space model of (\ref{eqn:Active_power_tf}) corresponding to a given $\epsilon_P$.  Given that the system is stable for each $\epsilon_P$, $A_P(\epsilon_P)$ is Hurwitz. Next, consider a solution to the Lyapunov equation $A_P^\top(\epsilon_P)W_P(\epsilon_P)+W_P(\epsilon_P)A_P(\epsilon_P)=-I$. Since $A_P(\epsilon_P)$ is Hurwitz for each $\epsilon_P\in[0,\epsilon_{max}]$, the eigenvalues of $W_P(\epsilon_P)$ are all strictly positive. Finally, let  $V_P(t)=x_P^\top(t)W_P(\epsilon_P(t))x_P(t)$, where $x_P$ is a state variable, be a Lyapunov candidate. 

If there exists a finite $\alpha>0$ such that $V_P(t)\leq \alpha V_P(0)$, then by Rayleigh Ritz theorem, 
\begin{align}
    \|x_P(t)\|_2^2\leq \alpha \frac{\max_{\epsilon_P\in[0,\epsilon_{max}]}\bar\lambda (W_P(\epsilon_P))}{\min_{\epsilon_P\in[0,\epsilon_{max}]}\underline\lambda (W_P(\epsilon_P))}\|x_P(0)\|_2^2.
\end{align}
Since the eigenvalues of $W_P(\epsilon)$ are strictly positive and $[0,\epsilon_{max}]$ is a closed interval, $\min_{\epsilon_P\in[0,\epsilon_{max}]}\underline\lambda(W_P(\epsilon_P))>0$, and the upper bound is finite.
Therefore, for any $\epsilon>0$, there exists $\delta>0$ such that if $\|x_0\|\leq \delta$, $\|x(t)\|\leq \epsilon$ and therefore, the system is stable.
\begin{proposition_1}
There exists a finite $\alpha>0$ during the transition.
\end{proposition_1}
{\bf Proof 1.} By taking the derivative of the Lyapunov candidate function and applying Rayleigh Ritz theorem, we obtain
\begin{eqnarray*} 
     \dot V_P(t)
     \leq \bar\lambda (\dot W_P(\epsilon_P(t))-I)\|x_P(t)\|_2^2
     \leq q(t)V_P(t),\ \mbox{where}
\end{eqnarray*}
 $$q(t)=\begin{cases}
      \frac{|\bar \lambda(\dot W_P(\epsilon_P(t)))|}{\min_{\epsilon_P\in[0,\epsilon_{max}]}\underline\lambda(W_P)} & \text{if $\dot \epsilon_P\neq0$}\\
      -\frac{1}{\max_{\epsilon_P\in[0,\epsilon_{max}]}\bar\lambda(W_P)} & \text{if $\dot \epsilon_P=0$}
    \end{cases}$$
Note that during the transition $\dot \epsilon_P\neq 0$, and $\bar\lambda (\dot W_P(\epsilon_P(t))-I)=\bar\lambda (\dot W_P(\epsilon_P(t)))-1\leq |\bar\lambda (\dot W_P(\epsilon_P(t)))|$. Conversely, when remaining in one mode, $\dot \epsilon_P=0$ and $\dot W_P(t)=0$ as $\dot W_P=\frac{\partial W_P}{\partial \epsilon_P}\dot\epsilon_P$. Then, $\bar\lambda (\dot W_P(\epsilon_P(t))-I)=-1$

By applying Gronwall-Bellman inequality, we obtain 
     $V_P(t)\leq V_P(0)exp\big(\int_{0}^t q(\tau)d\tau \big)$, and it is bounded since 
\begin{align*}
    \int_0^{t} q(\tau) d\tau 
    &\leq \frac{\max_{\epsilon_P\in[0,\epsilon_{max}]}\|\partial W_P(\epsilon_P)/\partial \epsilon_P\|}{\min_{\epsilon_P\in[0,\epsilon_{max}]}\underline\lambda(W_P)} \int_0^{t}  |\dot\epsilon_P(t)|d\tau\\
    &\leq\frac{\max_{\epsilon_P\in[0,\epsilon_{max}]}\|\partial W_P(\epsilon_P)/\partial \epsilon_P\|}{\min_{\epsilon_P\in[0,\epsilon_{max}]}\underline\lambda(W_P)} \epsilon_{max},
\end{align*}
where we choose (and assume) a transition strategy where $\epsilon_P$ changes monotonically with time.
\begin{proposition_1}
    $\max_{\epsilon_P\in[0,\epsilon_{max}]}\left\lVert\frac{\partial W_P(\epsilon_P)}{\partial \epsilon_P}\right\lVert$ is finite.
\end{proposition_1}
{\bf Proof 2.}
Since the coefficients of the characteristic polynomial  (\ref{eqn:Active_power_tf}) are affine with respect to $\epsilon_P$; therefore   $A_P(\epsilon)$ is also affine;  that is $A_P(\epsilon_P)=A_P(0)+\epsilon_PB_P$ for some $B_P$. Also $\frac{\partial }{\partial \epsilon_P}e^{A_P(\epsilon_P)\tau}=B_P\tau e^{A_P(\epsilon_P)\tau}$. 
    
If we choose $\epsilon_P(t)$ to be a smooth function of $t$, and using $W_P=\big(e^{A_P^\top(\epsilon_P)\tau}e^{A_P(\epsilon_P)\tau}\big)$, we get
\begin{eqnarray*}
    \left\lVert\frac{\partial W_P}{\partial \epsilon_P}\right\lVert\leq 2\|B_P\|\int_0^\infty  \tau \left\lVert e^{A_P^\top(\epsilon_P)\tau}e^{A_P(\epsilon_P)\tau}\right\lVert d\tau,&&
\end{eqnarray*}
where since $A_P(\epsilon_P)$ is Hurwitz, $\|e^{A_P^\top(\epsilon_P)\tau}e^{A_P(\epsilon_P)\tau}\|\leq e^{-\lambda_P \tau}$ for some $\lambda_P>0$. As a result, it is bounded for each $\epsilon_P$. Furthermore, since $[0,\epsilon_{max}]$ is a closed and bounded interval, $\max_{\epsilon_P\in[0,\epsilon_{max}]} \left\lVert \partial W_P/\partial \epsilon_P\right\lVert$ is bounded.

For multiple transitions, if there is a time gap of at least $\max_{\epsilon_P\in[0,\epsilon_{max}]}\bar\lambda(W_P)\frac{\max_{\epsilon_P\in[0,\epsilon_{max}]}\|\partial W_P(\epsilon_P)/\partial \epsilon_P\|}{\min_{\epsilon_P}\underline\lambda(W_P)} \epsilon_{max}$ between the end of one transition and the start of new transition, it ensures that $V_P(T)<V_P(0)$ where $T$ is the time when the new transition start. Then, the same upper bound for $V(t)$ can be applied to the new transition, ensuring stability.

Similar procedures can be followed for analyzing the transition stability of reactive power dynamics.

\section{Simulation Results}
{\bf Implementation:}
We designed the droop controller for the proposed GFL inverter under the assumption that we have a voltage controller with a large bandwidth. To design such a controller, let us first consider the dynamics of the LC filter of the inverter in Fig. \ref{fig:circuits}. In dq coordinates, the dynamics of the LC filter are given as follows:
\begin{align}
\begin{split}\label{eqn:il_dym}
&L_i\frac{d i_L^{d,q}}{dt}=m^{d,q} \frac{V_{dc}}{2}\pm L_i\omega i_L^{q,d}-V_c^{d,q}-R_i i_L^{d,q}
\end{split}
\end{align}

Based on \cite{yazdani2010voltage}, with the current controller $K_c$ and feedback linearizing control that cancels out the $d$ and $q$ coupling terms, the control input can be represented as follows
\begin{align}
    \begin{split}\label{eqn:Current_control}
        & \hat{m}^{d,q}=\frac{K_c(\hat{i}_{L,ref}^{d.q}-\hat{i}_L^{d.q})\mp L_i\omega_0 \hat{i}_L^{q,q}+\hat{V_c}^{d,q}}{V_{dc}/2}.
    \end{split}
\end{align}
Then, $\hat i_L^{d,q}=T_c \hat i_{L,ref}^{d,q}$
where $T_c=\frac{K_cG_i}{1+K_cG_i}$ and $G_i=\frac{1}{L_is+R_i}$.
On top of the current control in (\ref{eqn:Current_control}), by choosing $i_{L,ref}$ as
\begin{align}
\begin{split}
      \hat i_{L,ref}^{d,q}=K_v(\hat V_{c,ref}^{d,q}-\hat V_c^{d,q})\mp C_i \omega_0 \hat V_c^{q,d}+\hat i_g^{d,q},
\end{split}
\end{align}
one can achieve
$\hat V_c^{d,q}=T_v\hat V_{c,ref}^{d,q}$,
where $T_v=\frac{K_vTcG_v}{1+K_vTcG_v}$ and $G_v=\frac{1}{C_i s}$.

In this experiment, controllers with the following structure are utilized:
\begin{align}\label{eqn:implemented controller}
    \begin{split}
        K_c=\frac{L_is+R_i}{\tau_c s},\quad K_v=k_{p,V}+\frac{k_{i,V}}{s},
    \end{split}
\end{align}
where parameters $\tau_c$, $k_{p,V}$, and $k_{i,V}$ are tuned to have enough bandwidth.
Also, the conventional GFL that is used for comparison is using the same $K_c$. 

The proposed controller and the conventional GFL were implemented and tested using the MATLAB Simscape toolbox with a step size of 0.05 s.
The grid and inverter parameters used in the simulations are as follows: $R_i = 0.2\Omega$, $L_i = 3.3mH$, $C_i = 40\mu F$, $R_g = 0.1 \Omega$, $L_g = 1.86mH$. Additionally, the nominal voltage and frequency are set to $V_0 = 391V$ and $f_0 = 60Hz$. The chosen controller parameters are $k_{p,P} = 0.2 , \text{rad/(s}\cdot\text{ kW)}$, $k_{i,P}=0.1\omega_{lpf}\cdot k_{p,P}$, $k_{p,Q} = 0.2\text{ V/kVAR}$, and $k_{i,Q}=\omega_{lpf}\cdot k_{p,Q}$.

\begin{figure}[h]
\centering
\vspace{-5pt}
    \includegraphics[width=1.05\linewidth]{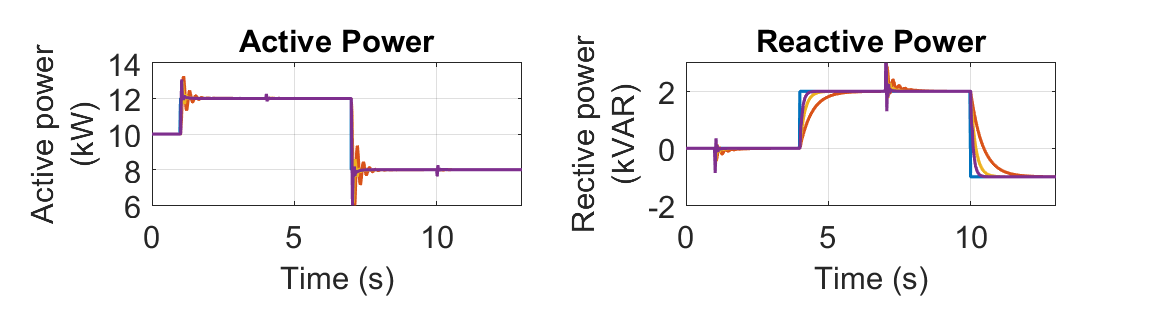}
    \caption{Power reference (blue line) tracking of proposed GFL (red, orange, and purple line correspond to $\omega_{lpf}$ values of $4\pi\text{ rad/s}$, $10\pi\text{ rad/s}$, and $20\pi\text{ rad/s}$ respectively)}
    \label{fig:Power_tracking}
    \vspace{-10pt}
\end{figure}

{\bf (a) Power Tracking:}
First, we evaluate the active and reactive power tracking of the proposed controller with filter cutoff frequencies ($\omega_{lpf}$) set to $4\pi \text{ rad/s}$, $5\pi\text{ rad/s}$, and $20\pi\text{ rad/s}$.
We assume that the inverter with the proposed controller is connected to a stiff grid with $\omega_g=\omega_0$ and $V_g=V_0$.  As shown in Fig.\ref{fig:Power_tracking}, at $t=1$ second, the active power reference is changed from 10kW to 12kW, and at $t=4$ second, the reactive power set point is changed from 0kVAR to 2kVAR. Similarly, at $t=7$ seconds, the active power $P_0$ changes from 12kW to 8kW, and at $t=10$ seconds, the reactive power $Q_0$ changes from 2kVAR to -1kVAR. The proposed controller achieves steady-state tracking just as conventional GFL.

\begin{figure}[h]
    \vspace{-10pt}
    \centering
    \includegraphics[width=1.05\linewidth]{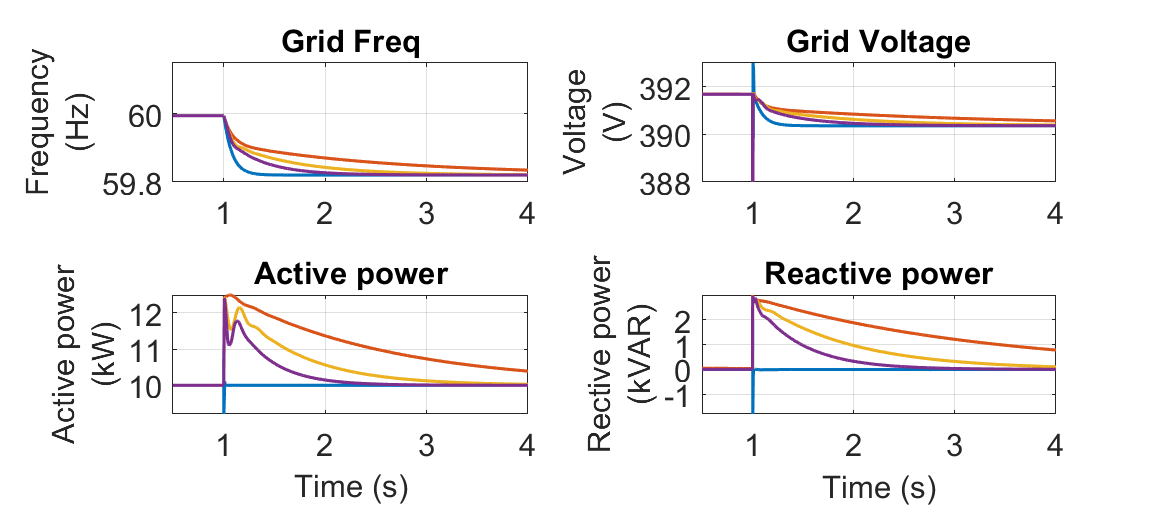}
    \caption{Response against sudden load change with Conventional (blue line) and Proposed GFL(red, orange, and purple line correspond to $\omega_{lpf}$ values of $4\pi\text{ rad/s}$, $10\pi\text{ rad/s}$, and $20\pi\text{ rad/s}$ respectively)}
    \label{fig:Grid_support_VF}
    \vspace{-10pt}
\end{figure}

{\bf (b) Response to Fast and Slow Load Changes:}
For checking the frequency and voltage support of the proposed GFL, consider the case where the GFL inverter is connected to the GFM inverter, forming a weak microgrid. Initially, there is a 20kW active load and a 10kVAR reactive load, and the GFL inverter injects 10kW of active power. Throughout this scenario, $P_0=10kW$ and $Q_0=0 VAR$. At $t=1$s, 5kW of active load and 5kVAR of reactive load are added to the grid. Then, the response is given as Fig.\ref{fig:Grid_support_VF}

For conventional GFL, active and reactive output power converge to the reference value immediately. Then, a GFM inverter solely handles load changes and experiences a huge rate of change of frequency (RoCoF). However, with the proposed controller, in the event of a load change, the GFL inverter temporarily supplies power, resulting in smaller voltage and frequency peaks and reduced RoCoF until the grid reaches new steady-state values. When conventional GFL shows a RoCoF of $1.74$ Hz/s, the proposed controllers with $\omega_{lpf}$ of $4\pi$, $10\pi$, and $20\pi$ achieve $0.89$ Hz/s, $0.93$ Hz/s, and $0.98$ Hz/s, respectively, which represent $51.35\%$, $53.30\%$, and $56.00\%$ of the RoCoF compared to the conventional GFL case. Additionally, the proposed GFL controller provides damping on the voltage.

\begin{figure}[h]
    \centering
    \vspace{-5pt}
    \includegraphics[width=1.05\linewidth]{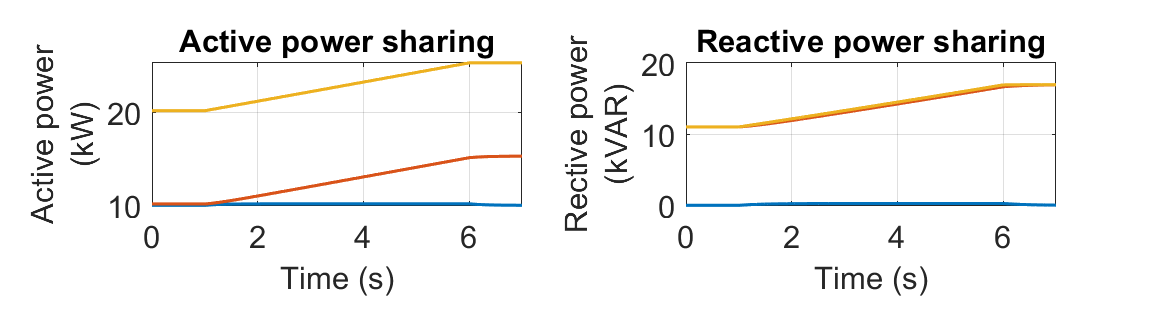}
    \caption{Power sharing between GFM and proposed GFL under slow load change: Blue represents the proposed GFL, Red represents GFM, and Orange represents the total load.}
    \label{fig:slow_load_change}
    \vspace{-10pt}
\end{figure}

Also, to verify that the proposed controller follows the power reference point and does not react to a slow load change, we apply 5 kW and 5 kVAR of active and reactive load changes over 5 seconds in a ramp manner to a microgrid with one GFM inverter and a proposed GFL inverter with $\omega_{lpf}=20\pi\text{ rad/s}$. As seen in Fig. \ref{fig:slow_load_change}, when the load changes slowly, most of the load change is handled by the GFM inverter, and the GFL inverter with the proposed controller maintains its power reference point.

\begin{figure}[h]
\vspace{-5pt}
\centering
    \includegraphics[width=1.05\linewidth]{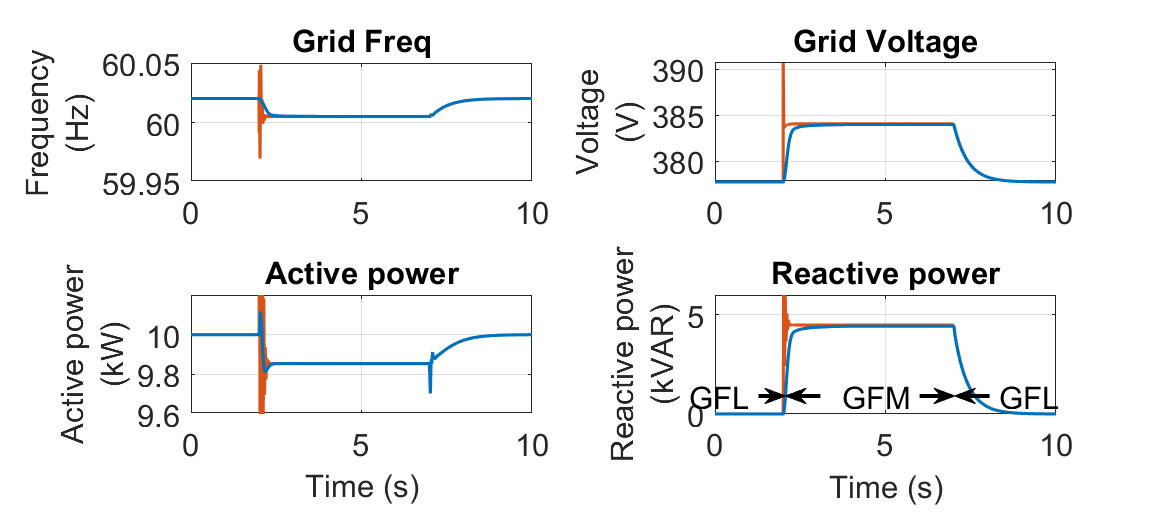}
    \caption{Transition response between GFL and GFM mode: Blue represents proposed transition, Red represents sudden transition.}
    \vspace{-15pt}
    \label{fig:Transition}
\end{figure}

{\bf (c) Transition between GFM and GFL:}
In the same microgrid setup with a GFL inverter utilizing the proposed controller ($\omega_{lpf}=20\pi\text{ rad/s}$) and a GFM inverter, a transition between GFL and GFM is tested (Fig. \ref{fig:Transition}). Initially, the proposed controller is in GFL mode. At $t=2$ s, the GFL inverter transitions to GFM mode. Parameters $\dot \epsilon=100$ and $\epsilon_{max}=200$ are chosen for smooth transition. For comparison, a sudden transition method proposed in \cite{chakraborty2023seamless} is also tested.
Both methods exhibit stability, as shown in active and reactive power plots (Fig. \ref{fig:Transition}). However, the smooth transition method demonstrates significantly lower overshoot and fluctuations compared to the sudden transition method. Specifically, during the transition, the sudden method results in a $4.1 kW$ active power overshoot, whereas the proposed method shows only $41W$. Similarly, the sudden method leads to a $4.4kVAR$ reactive power overshoot, while the proposed method experiences none.

Furthermore, a transition from the GFM mode to the GFL mode is tested at $t=7$ s. Here, for both methods, as explained before, $\epsilon_P$ and $\epsilon_Q$ jump from $\epsilon_{max}$ to 0. Even though the $\epsilon$ value changes abruptly from $\epsilon_{max}$ to 0, the transient is not significant because in GFL mode, control signals increases gradually and the transient is small.
\vspace{-5pt}
\section{CONCLUSIONS}
In this paper, we evaluate the robustness of conventional GFL and GFM inverters against power perturbations by analyzing their power-to-voltage responses. We observe that conventional GFL exhibits sensitivity to perturbations, whereas the droop control structure in GFM enables shaping of this sensitivity. Motivated by this, we propose a GFL control structure that is insensitive to high-frequency load perturbations and tracks low-frequency power references. 
Furthermore, drawing on the structural similarity between the proposed GFL controller and the GFM controller, we devise a smooth transition strategy between GFL and GFM modes. This strategy minimizes power overshoot and oscillation during the transition. 

\addtolength{\textheight}{-12cm}   









\bibliographystyle{Bibliography/IEEEtran}
\bibliography{Bibliography/ACC_2024_final_submission} 

\begin{thebibliography}{10}
\providecommand{\url}[1]{#1}
\csname url@rmstyle\endcsname
\providecommand{\newblock}{\relax}
\providecommand{\bibinfo}[2]{#2}
\providecommand\BIBentrySTDinterwordspacing{\spaceskip=0pt\relax}
\providecommand\BIBentryALTinterwordstretchfactor{4}
\providecommand\BIBentryALTinterwordspacing{\spaceskip=\fontdimen2\font plus
\BIBentryALTinterwordstretchfactor\fontdimen3\font minus \fontdimen4\font\relax}
\providecommand\BIBforeignlanguage[2]{{%
\expandafter\ifx\csname l@#1\endcsname\relax
\typeout{** WARNING: IEEEtran.bst: No hyphenation pattern has been}%
\typeout{** loaded for the language `#1'. Using the pattern for}%
\typeout{** the default language instead.}%
\else
\language=\csname l@#1\endcsname
\fi
#2}}

\bibitem{elkhatib2017evaluation}
M.~Elkhatib, J.~Neely, and J.~Johnson, ``Evaluation of fast-frequency support functions in high penetration isolated power systems,'' in \emph{2017 IEEE 44th Photovoltaic Specialist Conference (PVSC)}.\hskip 1em plus 0.5em minus 0.4em\relax IEEE, 2017, pp. 2141--2146.

\bibitem{qi2021synthetic}
Y.~Qi, H.~Deng, X.~Liu, and Y.~Tang, ``Synthetic inertia control of grid-connected inverter considering the synchronization dynamics,'' \emph{IEEE Transactions on Power Electronics}, vol.~37, no.~2, pp. 1411--1421, 2021.

\bibitem{wang2022study}
J.~Wang and G.~Saraswat, ``Study of inverter control strategies on the stability of low-inertia microgrid systems,'' in \emph{IECON 2022--48th Annual Conference of the IEEE Industrial Electronics Society}.\hskip 1em plus 0.5em minus 0.4em\relax IEEE, 2022, pp. 1--6.

\bibitem{chakraborty2023seamless}
S.~Chakraborty, S.~Patel, G.~Saraswat, A.~Maqsood, and M.~V. Salapaka, ``Seamless transition of critical infrastructures using droop controlled grid-forming inverters,'' \emph{IEEE Transactions on Industrial Electronics}, 2023.

\bibitem{yazdani2010voltage}
A.~Yazdani and R.~Iravani, \emph{Voltage-sourced converters in power systems: modeling, control, and applications}.\hskip 1em plus 0.5em minus 0.4em\relax John Wiley \& Sons, 2010.

\bibitem{askarian2024enhanced}
A.~Askarian, J.~Park, and S.~Salapaka, ``Enhanced grid following inverter (e-gfl): A unified control framework for stiff and weak grids,'' \emph{IEEE Transactions on Power Electronics}, 2024.

\bibitem{zhong2016universal}
Q.-C. Zhong and Y.~Zeng, ``Universal droop control of inverters with different types of output impedance,'' \emph{IEEE access}, vol.~4, pp. 702--712, 2016.

\bibitem{driesen2008virtual}
J.~Driesen and K.~Visscher, ``Virtual synchronous generators,'' in \emph{2008 IEEE power and energy society general meeting-conversion and delivery of electrical energy in the 21st century}.\hskip 1em plus 0.5em minus 0.4em\relax IEEE, 2008, pp. 1--3.

\bibitem{liu2015comparison}
J.~Liu, Y.~Miura, and T.~Ise, ``Comparison of dynamic characteristics between virtual synchronous generator and droop control in inverter-based distributed generators,'' \emph{IEEE Transactions on Power Electronics}, vol.~31, no.~5, pp. 3600--3611, 2015.

\bibitem{askarian2021control}
A.~Askarian, J.~Park, and S.~Salapaka, ``Control design for inverters: Beyond steady-state droop laws,'' \emph{arXiv preprint arXiv:2105.02292}, 2021.

\end{thebibliography}

\end{document}